\documentclass[12pt]{iopart}

\usepackage{graphicx}
\usepackage{iopams}

\begin{document}

\thispagestyle{empty}

\title{Nonperturbative theory of atom-surface interaction:
Corrections at short separations}

\author{M~Bordag,${}^{1}$ G~L~Klimchitskaya${}^{2,3}$ and V~M~Mostepanenko${}^{2,3,4}$
}

\address{${}^1$Institute for Theoretical
Physics, Leipzig University, Postfach 100920,
D-04009, Leipzig, Germany}
\address{${}^2$Central Astronomical Observatory at Pulkovo of the
Russian Academy of Sciences, Saint Petersburg,
196140, Russia}
\address{${}^3$Institute of Physics, Nanotechnology and
Telecommunications, Peter the Great Saint Petersburg
Polytechnic University, Saint Petersburg, 195251, Russia}
\address{${}^4$Kazan Federal University, Kazan, 420008, Russia}

\ead{vmostepa@gmail.com}

\begin{abstract}
The nonperturbative expressions for the free energy and force
of interaction between a ground-state atom and a real-material
surface at any temperature are presented. The
transition to the Matsubara representation is
performed, whereupon the comparison is made with the
commonly used
perturbative results based on the standard Lifshitz
theory. It is shown that the Lifshitz formulas for the
free energy and force of an atom-surface interaction follow
from the nonperturbative ones in the lowest order of the small
parameter. Numerical computations of the free energy
and force for the atoms of He${}^{\ast}$ and Na
interacting with a surface of an Au plate have been performed using
the frequency-dependent dielectric permittivity of Au and highly
accurate dynamic atomic polarizabilities in the framework of
both the nonperturbative and perturbative theories. According to
our results, the maximum deviations between the two
theories are reached at the shortest atom-surface
separations of about 1\,nm. Simple analytic expressions
for the atom-surface free energy are derived in the
classical limit and for an ideal-metal plane. In the lowest
order of the small parameter, they are
found in agreement with the perturbative ones following
from the standard Lifshitz theory. Possible applications
of the obtained results in the theory of van der Waals
adsorption are discussed.
\end{abstract}
\submitto{\JPCM}

\maketitle

\section{Introduction}

It is well known that an electrically neutral but
polarizable atom is attracted to a closely spaced material
surface by the force which increases rapidly with decreasing
atom-surface separation. The long-range forces of this kind
were predicted by van der Waals in the end of the
nineteenth century \cite{1}. After the development of
quantum field theory it was understood that the atom-surface
force is of entirely quantum nature and originates from
the zero-point and thermal fluctuations of the
electromagnetic field modified by a material boundary.
According to nonrelativistic London's theory \cite{2}, the
interaction potential between a polarizable atom and an
ideal-metal plane at zero temperature is inverse
proportional to the third power of their separation
\cite{3}.

The account of relativistic effects leads, however, to
important changes in the character of atom-surface interaction.
As was shown in the seminal work by Casimir and Polder
\cite{4}, at sufficiently large separations at zero
temperature the interaction potential between an atom and
an ideal-metal plane becomes inverse proportional to the
fourth power of separation. This result was obtained using
the second-order perturbation theory in the dipole-dipole
interaction within the framework of quantum electrodynamics.
The nonperturbative generalization of the
zero-temperature Casimir-Polder force between an atom and
an ideal-metal plane has been derived only a few years
ago \cite{5}.

One more phenomenon caused by the zero-point fluctuations
of the electromagnetic field is the attractive force
between two plane parallel, uncharged ideal-metal surfaces
in vacuum. The exact (nonperturbative) expression for this
force at zero temperature was obtained by Casimir \cite{6}
in the framework of quantum electrodynamics. The Lifshitz
theory \cite{7} presented a far-reaching generalization
of the Casimir result for the case of two plane parallel
material surfaces (semispaces) kept at any temperature $T$ in
thermal equilibrium with the environment. In so doing the
materials of the walls were described by the frequency-dependent
dielectric permittivities. The Lifshitz formulas for the
fluctuation-induced force contain the limiting case of
the nonrelativistic van der Waals force and, for the
ideal-metal planes, reproduce the Casimir result \cite{6}
as well as its generalization to nonzero temperature.

By rarefying the material of one of the semispaces, it is
customary to obtain the Lifshitz formulas for the free
energy and force between a polarizable atom and a surface
of the plate made of some real material at temperature
$T$ \cite{7,7a,8,9}.
The Lifshitz formulas describing an atom-surface interaction
are not as exact as the original Lifshitz formulas
derived for the case of two parallel surfaces. In fact the
former are obtained from the latter in the lowest
perturbation order of the small parameter equal to the
dynamic atomic polarizability multiplied by the number
of atoms per unit volume in the rarefied plate. If the
second wall is made of an ideal metal at zero temperature,
the Lifshitz formulas for an atom-surface interaction return
us back to the perturbative Casimir and Polder result
\cite{4}.

It should be particularly emphasized that the Lifshitz
formulas describing the atom-surface interaction have found
a lot of recent applications in various areas. They have
been used to investigate the dependences of atom-surface
forces on the material properties of a surface and on the
characteristics of an atom
\cite{10,11,12,13,14,15,16,17,18}.
The obtained information was used for
interpretation of experiments on quantum reflection
\cite{19,20,21}, Bose-Einstein condensation
\cite{22,23,24,25}, investigation of the resonance interaction
of two atoms near an ideal-metal wall \cite{26a}, and of the
role played by fluctuations at the interface between a metallic
boundary and the vacuum \cite{26b}.
The extent of agreement between the
measured atom-surface forces and the Lifshitz theory has
been also used in fundamental physics to place stronger
limits on the Yukawa-type corrections to Newton's law
of gravitation \cite{26} and on the parameters of
axion-like particles \cite{27}. Currently the Lifshitz
formulas are employed to describe the interaction of
different atoms with graphene and graphene-coated
substrates \cite{28,29,30,31,31a,32}. All the above makes
important the derivation of the nonperturbative
Lifshitz-type formula describing the free energy of an
atom interacting with a surface of the plate made of real
material at any nonzero temperature. In the framework of
the Lifshitz theory this formula would be as exact as the
famous Lifshitz expression describing the Casimir interaction
between two parallel material plates.

The recently developed formalism \cite{32a} allows a
unified nonperturbative description of the
fluctuation-induced forces between two atoms, an atom
and a material surface (plate), and between two material
plates with taken into account dissipation properties
at any nonzero temperature. This formalism is based
on the heat bath approach and earlier works devoted to
a quantum oscillator interacting with a blackbody
radiation field \cite{32b} and to nonperturbative
derivation of the van der Waals \cite{32c} and
Casimir-Polder \cite{32d} interaction between two
oscillating dipoles. For the case of two material
plates paper \cite{32a} reproduces the familiar Lifshitz
formula, whereas for an atom interacting with a
material surface
suggests its nonperturbative generalization.

In this paper, we compare the nonperturbative formulas for the
free energy and force of a ground-state atom,
interacting with a surface of
thick real-material plate (semispace)
at temperature $T$, and the commonly used perturbative Lifshitz
formulas obtained by means of the rarefying procedure.
In so doing, the atom is characterized by the dynamic
atomic polarizability and the plate material by the
frequency-dependent dielectric permittivity. We show
that the commonly known Lifshitz formulas for the free
energy and force of an atom-surface interaction are
reproduced from the exact ones as the first expansion
orders in powers of the dynamic atomic polarizability
divided by the third power of separation. We perform
numerical computations of the free energy and force
for atoms of metastable helium He${}^{\ast}$ and Na interacting
with an Au plate by using the exact formulas and the
standard perturbative Lifshitz formulas. It is shown
that at the shortest separation distance, where the
Lifshitz theory is still applicable, the relative
deviations between the obtained free energies exceed
1.1\% for He${}^{\ast}$ and 0.6\% for Na. The respective
deviations in the force are 2.3\% and 1.25\% for the
atoms of He${}^{\ast}$ and Na, respectively. At separations
exceeding 4\,nm the exact and perturbative theories
lead to almost coinciding results. We also obtain
simple analytic expressions for the nonperturbative
atom-surface
free energies and forces in the classical limit and
for an ideal-metal plane and demonstrate that in these
cases the deviations between the exact and
perturbative results are negligibly small.

The paper is organized as follows. In section~2, we
briefly summarize the main results of the nonperturbative
theory of atom-surface interaction and make a transition
to the Matsubara representation. In section~3, we
reobtain the standard Lifshitz formulas
for an atom-surface interaction
in the first
perturbation order and perform numerical computations
of the free energy and force between metastable He${}^{\ast}$
and Na atoms
and an Au plate using both the exact and perturbative
formulas. Section~4 contains simple analytic results
in the classical limit and for an ideal-metal surface.
In section~5 the reader will find our conclusions and a
discussion.

\section{Nonperturbative formulas for the free energy and force of
atom-surface interaction}

We consider the polarizable atom in the ground state
described by the dynamic atomic polarizability
$\alpha(\omega)$ separated by a distance $a$ from the
surface of a
thick plate (semispace) kept at temperature $T$ in
thermal equilibrium with the environment. The
material of the plate is characterized by the
frequency-dependent dielectric permittivity
$\varepsilon(\omega)$. In the nonperturbative
theory, developed recently \cite{32a}, both the
external atom and the atoms of the plate are
considered as point dipoles interacting through the
electromagnetic field. The dynamics of the atoms
is described by the oscillator equations with some
mass, the intrinsic frequency and the damping
parameter to account for dissipation. The oscillators
are subjected to the external forces equal to the
sums of the electric and Langevin forces. The latter
may be introduced by coupling the heat baths to the
oscillators and integrating out the bath variables
\cite{32e}. To describe the wall, an infinite
collection of the dipoles, which are distributed
homogeneously in the half space, is taken with
subsequent increase of their density to form a
continuous distribution.

Finally, one arrives to the following expression for
the free energy of an atom-surface interaction \cite{32a}
\begin{equation}
{\cal F}(a,T)=\frac{1}{\pi}\!\int_0^{\infty}\!\!\!d\omega\left[
\frac{\hbar\omega}{2}+k_BT\ln\left(1-e^{-\frac{\hbar\omega}{k_BT}}
\right)\right]
\frac{\partial\Delta(\omega)}{\partial\omega},
\label{eq1}
\end{equation}
\noindent
where $k_B$ is the Boltzmann constant and the so-called {\it phase}
is defined as
\begin{equation}
\Delta(\omega)=-\frac{1}{2i}\ln\frac{L(\omega)}{L(-\omega)}.
\label{eq2}
\end{equation}
\noindent
The quantity ${L(\omega)}$ in (\ref{eq2}) can be expressed via the difference
$\Delta G$ of photon Green's functions in the presence of a material half-space
and of the free space
\begin{equation}
L(\omega)=1-4\pi\alpha(\omega)\frac{\omega^2}{c^2}\Delta G(\omega).
\label{eq3}
\end{equation}
\noindent
The latter difference is given by \cite{32a}
\begin{eqnarray}
&&
\Delta G(\omega)=\frac{1}{4\pi\omega^2}\int_0^{\infty}\frac{k_{\bot}dk_{\bot}}{q}
e^{-2aq}
\label{eq4} \\
&&~~
\times
\left[(2k_{\bot}^2c^2-\omega^2)r_{\rm TM}(\omega,k_{\bot})+
\omega^2r_{\rm TE}(\omega,k_{\bot})\right],
\nonumber
\end{eqnarray}
\noindent
where $k_{\bot}$ is the magnitude of the wave vector projection on the surface
of the plate, $q^2=k_{\bot}^2-\omega^2/c^2$, and the reflection coefficients for
the two independent polarizations of the electromagnetic field on this surface,
transverse magnetic (TM) and transverse electric (TE), take the usual form
\begin{eqnarray}
&&
r_{\rm TM}(\omega,k_{\bot})=\frac{\varepsilon(\omega)q-k}{\varepsilon(\omega)q+k},
\label{eq5} \\
&&
r_{\rm TE}(\omega,k_{\bot})=\frac{q-k}{q+k}, \quad
k^2\equiv k_{\bot}^2-\varepsilon(\omega)\frac{\omega^2}{c^2}.
\nonumber
\end{eqnarray}

It is convenient now to pass in (\ref{eq1}) to the Matsubara  representation.
For this purpose we present (\ref{eq2}) for the phase as
\begin{equation}
\Delta(\omega)=-\frac{1}{2i}\ln{L(\omega)}+\frac{1}{2i}\ln{L(-\omega)}.
\label{eq6}
\end{equation}
\noindent
We take into account that in the first term on the right-hand side of (\ref{eq6})
all singularities are in the lower half of the plane of complex frequency, whereas
the singularities of the second term are in the upper half-plane.
When substituting (\ref{eq6}) in (\ref{eq1}), we turn the integration path in
the first term of the free energy towards the positive imaginary axis, $\omega=i\xi$,
and in the second term --- to the negative imaginary axis.

The expression in the square brackets in (\ref{eq1}) can be written in the form
\begin{equation}
\frac{\hbar\omega}{2}+k_BT\ln(1-e^{-\hbar\omega/(k_BT)})=
k_BT\ln\left(2\sinh\frac{\hbar\omega}{2k_BT}\right).
\label{eq7}
\end{equation}
\noindent
The analytic continuation of the logarithm on the right-hand side of (\ref{eq7})
is performed according to
\begin{eqnarray}
&&
\ln\left(2\sinh\frac{i\hbar\xi}{2k_BT}\right)=
\ln\left|2\sin\frac{\hbar\xi}{2k_BT}\right|+
i\pi\sum_{l=0}^{\infty}{\vphantom{\sum}}^{\prime}
\theta(\xi-\xi_l),
\nonumber \\[-2mm]
&&\label{eq8}\\[-2mm]
&&
\ln\left(2\sinh\frac{i\hbar\xi}{2k_BT}\right)=
\ln\left|2\sin\frac{\hbar\xi}{2k_BT}\right|-
i\pi\!\sum_{l=-\infty}^{0}{\vphantom{\sum}}^{\!\!\!\prime}
\theta(-\xi+\xi_l)
\nonumber
\end{eqnarray}
\noindent for $\xi>0$ and $\xi<0$, respectively.
Here $\xi_l=2\pi k_BTl/\hbar$, $l=0,\,\pm1,\,\pm2,\,\ldots\,$
are the Matsubara frequencies, $\theta(x)$ is the Heaviside step function,
and the prime on the summation sign denotes that the contribution from the
zeroth term should be taken with the factor 1/2.
The relation (\ref{eq8}) follows from the fact that the logarithm has cuts
starting at $\xi=\xi_l$.

Substituting (\ref{eq6}) and (\ref{eq7}) in (\ref{eq1})
and turning the integration paths as described above with account of (\ref{eq8}), we find
\begin{equation}
{\cal F}(a,T)=-k_BT\int_{0}^{\infty}\!\!d\xi\sum_{l=0}^{\infty}{\vphantom{\sum}}^{\prime}
\theta(\xi-\xi_l)\frac{\partial}{\partial\xi}L(i\xi).
\label{eq9}
\end{equation}
\noindent
After integrating by parts in (\ref{eq9}), we come to the final formula
for the free energy of atom-surface interaction
\begin{equation}
{\cal F}(a,T)=k_BT\sum_{l=0}^{\infty}{\vphantom{\sum}}^{\prime}
\ln L(i\xi_l).
\label{eq10}
\end{equation}
\noindent
For the function $L(i\xi_l)$ from (\ref{eq3}) and (\ref{eq4}) one obtains
\begin{eqnarray}
&&
L(i\xi_l)=1-\alpha(i\xi_l)\int_{0}^{\infty}\frac{k_{\bot}dk_{\bot}}{q_l}
e^{-2aq_l}
\label{eq11} \\
&&~
\times
\left[\left(2k_{\bot}^2+\frac{\xi_l^2}{c^2}\right)r_{\rm TM}(i\xi_l,k_{\bot})-
\frac{\xi_l^2}{c^2}r_{\rm TE}(i\xi_l,k_{\bot})\right],
\nonumber
\end{eqnarray}
\noindent
where  $q_l^2=k_{\bot}^2+\xi^2/c^2$ and $r_{\rm TM}$ and $r_{\rm TE}$ are again
given by (\ref{eq5}) with the substitution $\omega=i\xi_l$.

Calculating the negative derivative of the free energy (\ref{eq10}) and
(\ref{eq11}) with respect to separation, one obtains the nonperturbative expression
for the atom-surface force
\begin{eqnarray}
&&
F(a,T)=-2k_BT\sum_{l=0}^{\infty}{\vphantom{\sum}}^{\prime}
\frac{\alpha(i\xi_l)}{L(i\xi_l)}\int_{0}^{\infty}\!\!{k_{\bot}dk_{\bot}}
e^{-2aq_l}
\label{eq12} \\
&&~
\times
\left[\left(2k_{\bot}^2+\frac{\xi_l^2}{c^2}\right)r_{\rm TM}(i\xi_l,k_{\bot})-
\frac{\xi_l^2}{c^2}r_{\rm TE}(i\xi_l,k_{\bot})\right].
\nonumber
\end{eqnarray}

In the next section we compare the exact (nonperturbative)
results (\ref{eq10})--(\ref{eq12})
with the commonly used perturbative free energies and forces given by the
Lifshitz formulas for an atom-surface interaction.

\section{Computations of the atom-surface interaction by means of the
nonperturbative and perturbative theories}

It is convenient to rewrite the obtained nonperturbative expressions for the
free energy (\ref{eq10}) and force (\ref{eq12}) of atom-surface interaction in
terms of the dimensionless variables
\begin{equation}
y=2aq_l,\quad
\zeta_l=\frac{\xi_l}{\omega_c}=\frac{2a\xi_l}{c}.
\label{eq13}
\end{equation}
\noindent
Then the exact free energy (\ref{eq10}) takes the form
\begin{eqnarray}
&&
{\cal F}(a,T)=k_BT\sum_{l=0}^{\infty}{\vphantom{\sum}}^{\prime}
\ln \left\{1-\frac{\alpha(i\omega_c\zeta_l)}{8a^3}
\int_{\zeta_l}^{\infty}\!\!\!dy\,e^{-y}
\right.
\nonumber \\
&&~
\left.
\vphantom{\int_{\zeta_l}^{\infty}}
\times\left[(2y^2-\zeta_l^2)r_{\rm TM}(i\zeta_l,y)-
\zeta_l^2r_{\rm TE}(i\zeta_l,y)\right]\right\}.
\label{eq14}
\end{eqnarray}
\noindent
In a similar way, the exact atom-surface force (\ref{eq12})
with account of (\ref{eq11}) is given by
\begin{eqnarray}
&&
{F}(a,T)=-\frac{k_BT}{a}\sum_{l=0}^{\infty}{\vphantom{\sum}}^{\prime}
\frac{\alpha(i\omega_c\zeta_l)}{8a^3}
\label{eq15} \\
&&{\!\!\!\!\!\!\!\!\!\!}
\times
\frac{\int_{\zeta_l}^{\infty}dy\,ye^{-y}
\left[(2y^2-\zeta_l^2)r_{\rm TM}(i\zeta_l,y)-
\zeta_l^2r_{\rm TE}(i\zeta_l,y)\right]}{1-
\frac{\alpha(i\omega_c\zeta_l)}{8a^3}
\int_{\zeta_l}^{\infty}\!\!dy\,e^{-y}
\left[(2y^2-\zeta_l^2)r_{\rm TM}(i\zeta_l,y)-
\zeta_l^2r_{\rm TE}(i\zeta_l,y)\right]}.
\nonumber
\end{eqnarray}
\noindent

Expanding (\ref{eq14}) and (\ref{eq15}) up to the first power of
a small parameter $\alpha/a^3$, one obtains the standard, perturbative,
expressions \cite{7a,8,9,10}
for the free energy and force of an atom-surface interaction
\begin{eqnarray}
&&
{\cal F}^{\rm per}(a,T)=-\frac{k_BT}{8a^3}\sum_{l=0}^{\infty}{\vphantom{\sum}}^{\prime}
\alpha(i\omega_c\zeta_l)
\int_{\zeta_l}^{\infty}\!\!dy\,e^{-y}
\nonumber \\
&&~~
\times
\left[(2y^2-\zeta_l^2)r_{\rm TM}(i\zeta_l,y)-
\zeta_l^2r_{\rm TE}(i\zeta_l,y)\right],
\nonumber\\[-2mm]
&&\label{eq16}\\[-2mm]
&&
{F}^{\rm per}(a,T)=-\frac{k_BT}{8a^4}\sum_{l=0}^{\infty}{\vphantom{\sum}}^{\prime}
\alpha(i\omega_c\zeta_l)
\int_{\zeta_l}^{\infty}\!\!dy\,ye^{-y}
\nonumber \\
&&~~
\times
\left[(2y^2-\zeta_l^2)r_{\rm TM}(i\zeta_l,y)-
\zeta_l^2r_{\rm TE}(i\zeta_l,y)\right].
\nonumber
\end{eqnarray}

The reflection coefficients entering (\ref{eq14})--(\ref{eq16}) are
obtained from (\ref{eq5}) by the substitution $\omega=i\xi_l$ and
the change
of variables (\ref{eq13}):
\begin{eqnarray}
&&
r_{\rm TM}(i\zeta_l,y)=\frac{\varepsilon_ly-
\sqrt{y^2+\zeta_l^2(\varepsilon_l-1)}}{\varepsilon_ly+
\sqrt{y^2+\zeta_l^2(\varepsilon_l-1)}},
\nonumber \\
&&
r_{\rm TE}(i\zeta_l,y)=\frac{y-
\sqrt{y^2+\zeta_l^2(\varepsilon_l-1)}}{y+
\sqrt{y^2+\zeta_l^2(\varepsilon_l-1)}},
\label{eq17}
\end{eqnarray}
\noindent
where $\varepsilon_l\equiv\varepsilon(i\omega_c\zeta_l)$.
Note that (\ref{eq16}) and (\ref{eq17}) are well known as the Lifshitz
formulas for an atom-surface interaction \cite{7,7a,8,9}.

We calculate the free energies ${\cal F}$, ${\cal F}^{\rm per}$ and forces
${F}$, ${F}^{\rm per}$ of an atom-surface interaction for the atoms of metastable
helium, He${}^{\ast}$, and natrium, Na, spaced in close proximity to the Au plate.
Taking into account that the atomic polarizability $\alpha$ along the imaginary
frequency axis takes the maximum values of order of $10^{-29}\,\mbox{m}^3$, one
concludes that the dimensionless parameter $\alpha/a^3$, entering
(\ref{eq14}) and (\ref{eq15}), remains small down to $a=1\,$nm
separation. Since the description of a plate material by means of the dielectric
permittivity is applicable only at the distances much larger than the atomic sizes,
within the Lifshitz theory one cannot consider atom-surface separations smaller
than 8--10\,\AA. This means that the maximum deviations between the atom-surface
interactions computed exactly and perturbatively are expected at separations of
order of 1\,nm.

\begin{figure}[b]
\vspace*{-8cm}
\centerline{\hspace*{2.5cm}
\includegraphics{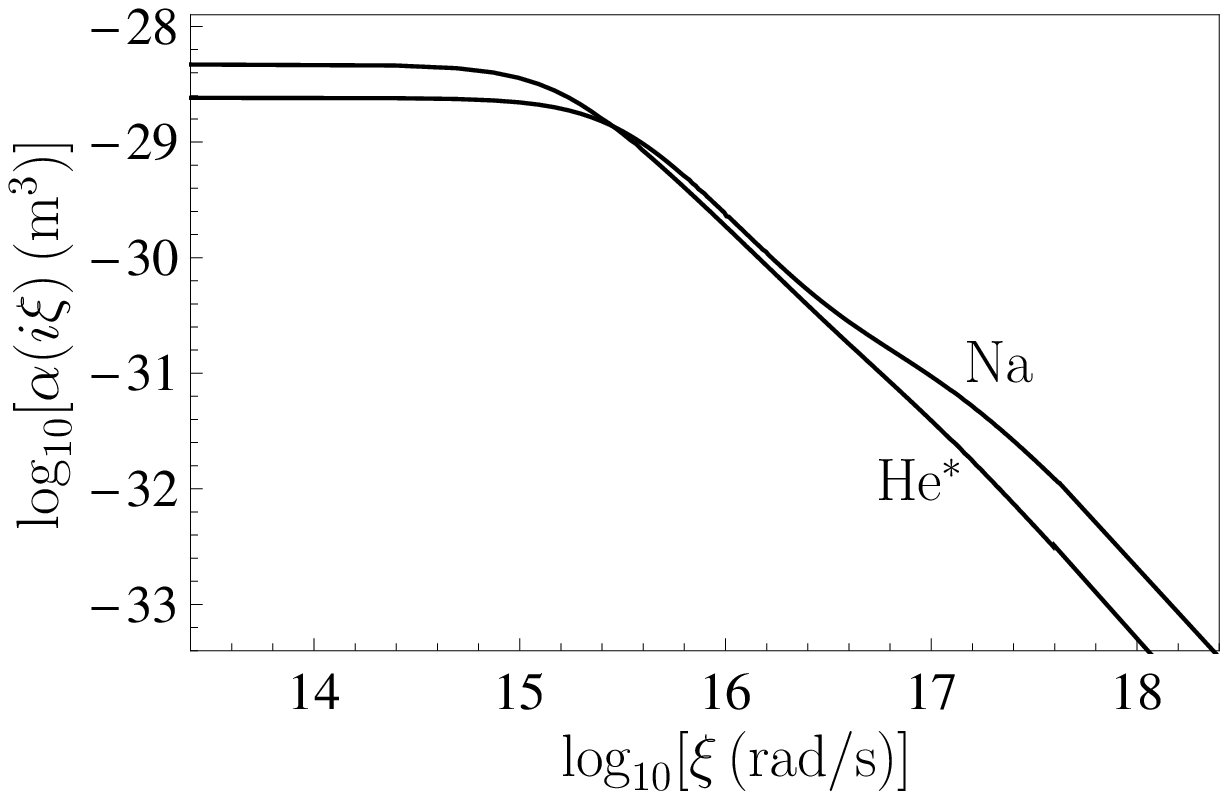}
}
\vspace*{-10cm}
\caption{\label{fg1}
The highly accurate dynamic atomic polarizabilities for the atoms
of He${}^{\ast}$ and Na are shown as functions of imaginary frequency
in the double logarithmic scale.
}
\end{figure}

Precise computations of the atom-surface free energy and force at separations of a few
nanometers require the knowledge of dynamic atomic polarizabilities accurate up to
the frequencies of order of $10^{17}-10^{18}\,$rad/s (the single-oscillator model
cannot be used at so short separation distances \cite{9,10,12}).
The nonrelativistic
highly accurate dynamic atomic polarizability of
He${}^{\ast}(2\,{}^{3\!}S)$  was obtained in \cite{33}  (see also \cite{34})
with the relative error of about $10^{-6}$.
It is shown by the line labeled He${}^{\ast}$ in figure~\ref{fg1} as a function of
frequency.
Note that in the relativistic framework the polarizability of He${}^{\ast}$
includes also the tensor part.
The static atomic polarizability of He${}^{\ast}$ is equal to
$\alpha_{{\rm He}^{\ast}}(0)=467.727\times 10^{-31}\,\mbox{m}^3$.
The highly accurate dynamic atomic polarizability of Na
in the ground state can be found in \cite{35}.
It is shown as a function of frequency by the line labeled Na in figure~\ref{fg1}
(for Na the static atomic polarizability  is equal to
$\alpha_{{\rm Na}}(0)=241.067\times 10^{-31}\,\mbox{m}^3$).

The dielectric permittivity of Au wall along the imaginary frequency axis enters
(\ref{eq14})--(\ref{eq16}) through the reflection coefficients (\ref{eq17}).
It is found from the optical data for the complex index of refraction of Au
\cite{36} using the Kramers-Kronig relation.
Since the optical data are available only within some restricted frequency region,
at low frequencies down to zero frequency they have to be extrapolated with the help
of some theoretical model. There is widely discussed problem \cite{9,37} that
an extrapolation by means of the Drude model, which takes into account the dissipation
of free electrons, results in drastic contradictions between theoretical predictions
of the Lifshitz theory and all precise experiments on measuring the Casimir force
for two metallic test bodies \cite{38,39,40,41,42,43,44,45}.
If the extrapolation is made by means of the lossless plasma model,
the experimental data are found to be in a very good agreement with theory
\cite{38,39,40,41,42,43,44,45,46}.
Different theoretical predictions arise due to different values taken by the
transverse electric reflection coefficient at zero frequency when two dissimilar
extrapolations of the optical data are employed.
Similar problem arises with the transverse magnetic reflection coefficient in the
case of dielectric surfaces when the free charge carriers are either taken into
account or omitted at nonzero temperature \cite{47,48,49}.
It should be stressed, however,
that for an atom-surface configuration under consideration here the Drude-plasma
dilemma does not influence the computational results. The reason is that in
(\ref{eq14}) and (\ref{eq15}), as well as in  (\ref{eq16}),
the transverse electric coefficient $r_{\rm TE}(0,y)$ does not contribute to
the result, as it is multiplied by the factor $\zeta_0^2=0$. To be specific,
in computations below we extrapolate the optical data for Au to zero frequency
by means of the experimentally consistent plasma model \cite{9,37}.

\begin{figure}[b]
\vspace*{-0cm}
\centerline{\hspace*{2.5cm}
\includegraphics{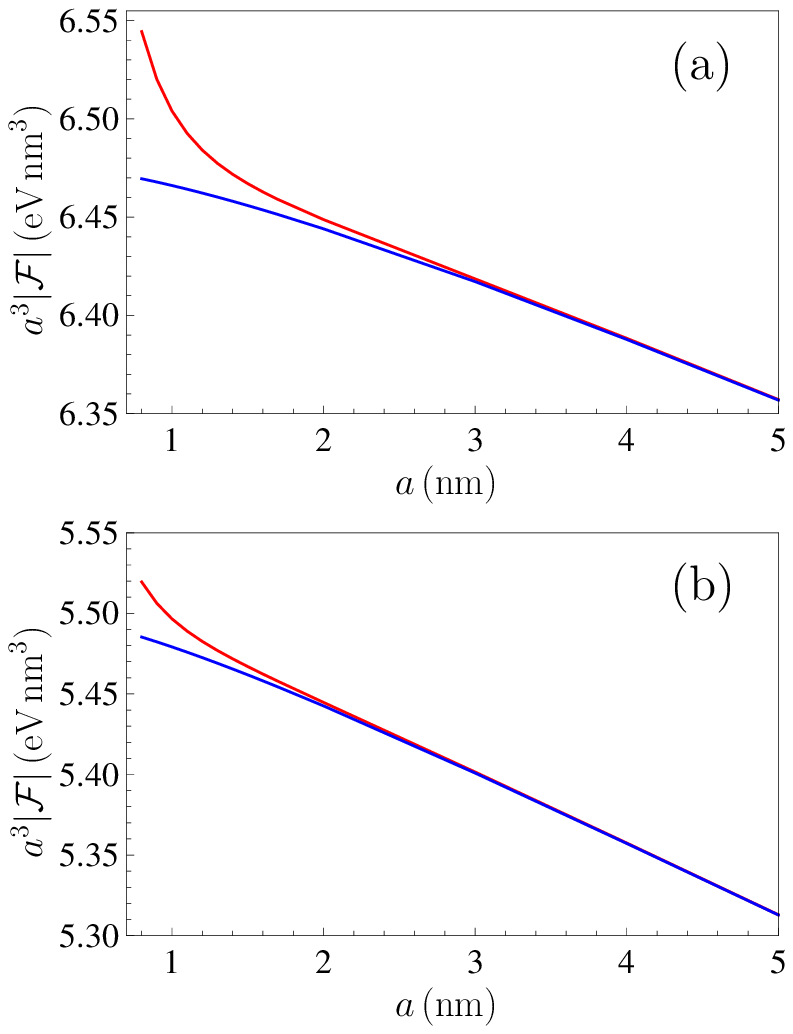}
}
\vspace*{-17cm}
\caption{\label{fg2}
The computational results for the magnitude of the free energy of
atom-surface interaction at $T=300\,$K multiplied by the third power
of separation are shown as functions of separation by the bottom and
top lines computed using the perturbative and exact theories,
respectively, for (a) the atom of He${}^{\ast}$ and
(b) the atom of Na.
}
\end{figure}

The magnitude of the (negative) free energy of an atom-surface interaction was computed
using the new expression (\ref{eq14}) and the standard, perturbative,
expression (\ref{eq16}).
The computational results multiplied by the third power of separation between
an atom and a wall are shown in figure~\ref{fg2} as functions of separation by the
bottom and top lines for the standard and exact theories, respectively.
Figure~\ref{fg2}(a) refers to the atom of metastable He${}^{\ast}$ and
figure~\ref{fg2}(b) --- to the atom of Na. Computations are performed at room
temperature $T=300\,$K. As is seen in figure~\ref{fg2}(a), the largest difference
between the exact and perturbative free energies for an atom of He${}^{\ast}$
is reached at the shortest separation considered, $a=0.8\,$nm, where it is equal
to approximately 0.146\,eV. For the atom of Na at $a=0.8\,$nm from an Au plate
the difference between the exact and perturbative predictions is equal to 0.053\,eV.
At $a>4\,$nm the exact and perturbative theories lead to almost coinciding
free energies and forces.

\begin{figure}[b]
\vspace*{-0cm}
\centerline{\hspace*{2.5cm}
\includegraphics{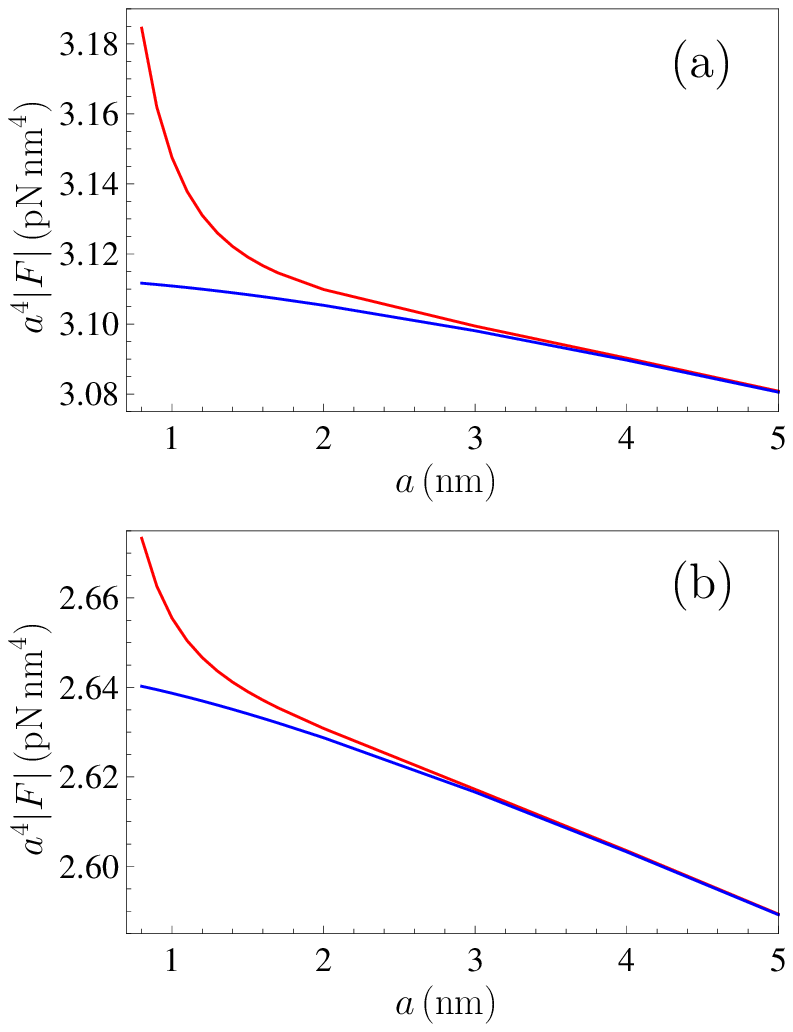}
}
\vspace*{-17cm}
\caption{\label{fg3}
The computational results for the magnitude of the force of
atom-surface interaction at $T=300\,$K multiplied by the fourth power
of separation are shown as functions of separation by the bottom and
top lines computed using the perturbative and exact theories,
respectively, for (a) the atom of He${}^{\ast}$ and
(b) the atom of Na.
}
\end{figure}

In figure~\ref{fg3} the computational results for the magnitude of atom-surface force
at $T=300\,$K multiplied by the fourth power of separation are plotted as the
bottom and top lines for the standard and exact theories, respectively.
Computations are performed by (\ref{eq15}) and (\ref{eq16}).
The case of an atom of He${}^{\ast}$ is shown in figure~\ref{fg3}(a) and
of an atom of Na --- in figure~\ref{fg3}(b).
The largest differences between theoretical predictions of the exact and
perturbative theories are again reached at the shortest separation considered
$a=0.8\,$nm. For the  atoms of He${}^{\ast}$ and Na the largest force
differences are equal to 0.178\,pN and 0.081\,pN, respectively.
Taking into account that at the moment the experimental sensitivity to small
forces is of the order of a fraction of 1\,fN \cite{45,50}, the above differences
between the exact and perturbative theories should be considered as quite
measurable.

\begin{figure}[b]
\vspace*{-0cm}
\centerline{\hspace*{2.5cm}
\includegraphics{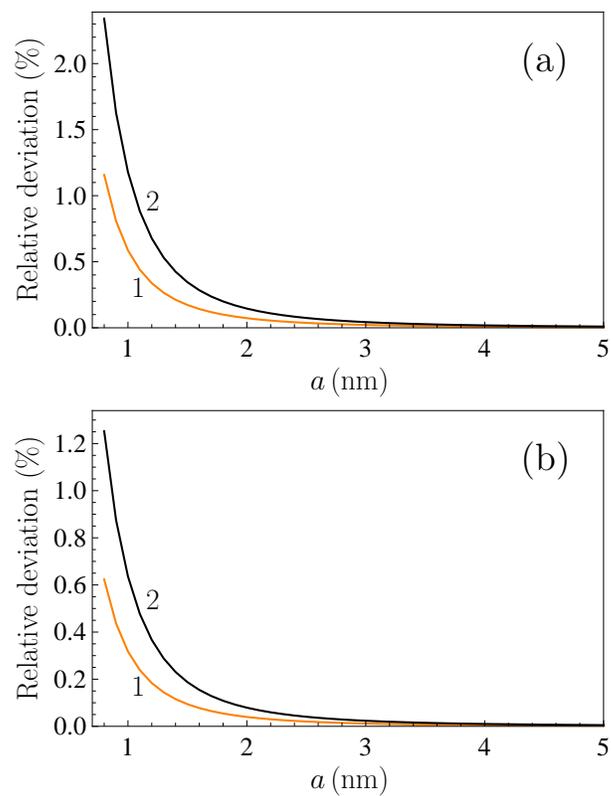}
}
\vspace*{-17cm}
\caption{\label{fg4}
The computational results for the relative deviations between the exact
and perturbative  free energies (lines labeled 1) and forces
(lines labeled 2) of atom-surface interaction at $T=300\,$K
are shown as functions of separation
for (a) the atom of He${}^{\ast}$ and
(b) the atom of Na.
}
\end{figure}

It is interesting also to consider the relative deviations between the free
energies and forces of an atom-surface interaction computed exactly and
perturbatively
\begin{eqnarray}
&&
\delta{\cal F}(a,T)=
\frac{{\cal F}(a,T)-{\cal F}^{\rm per}(a,T)}{{\cal F}^{\rm per}(a,T)},
\nonumber \\
&&
\delta{ F}(a,T)=
\frac{{ F}(a,T)-{ F}^{\rm per}(a,T)}{{ F}^{\rm per}(a,T)}.
\label{eq18}
\end{eqnarray}
\noindent
In figure~\ref{fg4} the relative deviations between the atom-surface free energies
(lines labeled 1) and forces (lines labeled 2) are shown at $T=300\,$K
as functions of separation for (a) the atom of He${}^{\ast}$ and (b) the atom
of Na. As is seen in figure~\ref{fg4}(a), for the atom of He${}^{\ast}$ the
maximum relative deviations in the free energy and force exceed  1.1\% and
2.3\%, respectively. For the atom of Na the maximum relative deviations are
0.6\% for the free energy and 1.25\% for the force.
At the atom-surface separations exceeding 4\,nm both the exact and perturbative
theories lead to almost coinciding predictions for free energies and forces.
Note that the decrease of temperature down to $T=100\,$K makes only a minor
impact on these results.

\section{Analytic results in classical limit and for ideal-metal surface}

It has been known that in the classical limit (i.e., at large separations or
high temperatures) the main contribution to the Lifshits formulas for the
free energy and force of an atom-surface interaction is given by the zero-frequency
term. The contribution of all terms with $l\geq 1$ remains exponentially
small in this case. It is easily seen that the same holds for the nonperturbative
atom-surface free energy (\ref{eq14}) and force (\ref{eq15}).
As an example, let us consider the contribution of the first Matsubara
 frequency to the free energy (\ref{eq14}), i.e.,
\begin{eqnarray}
&&
{\cal F}_1(a,T)=k_BT
\ln \left\{1-\frac{\alpha(i\omega_c\zeta_1)}{8a^3}
\int_{\zeta_1}^{\infty}dy\,e^{-y}
\right.
\label{eq19} \\
&&~~
\left.
\vphantom{\int_{\zeta_l}^{\infty}}
\times\left[(2y^2-\zeta_1^2)r_{\rm TM}(i\zeta_1,y)-
\zeta_1^2r_{\rm TE}(i\zeta_1,y)\right]\right\},
\nonumber
\end{eqnarray}

{}From (\ref{eq19}) one easily obtains
\begin{eqnarray}
&&
|{\cal F}_1(a,T)|\approx k_BT
\frac{\alpha(i\omega_c\zeta_1)}{8a^3}
\int_{\zeta_1}^{\infty}dy\,e^{-y}
\label{eq20} \\
&&~~
\times\left|2y^2r_{\rm TM}(i\zeta_1,y)-
\zeta_1^2[r_{\rm TM}(i\zeta_1,y)+r_{\rm TE}(i\zeta_1,y)]\right|.
\nonumber
\end{eqnarray}

Taking into account that the dynamic atomic polarizability decreases
with increasing frequency and that $|r_{\rm TE}|<r_{\rm TM}\leq 1$,
we find from (\ref{eq20})
\begin{eqnarray}
&&
|{\cal F}_1(a,T)|<k_BT
\frac{\alpha(0)}{4a^3}r_{\rm TM}(0)
\int_{\zeta_1}^{\infty}y^2e^{-y}dy
\label{eq21} \\
&&~~
=k_BT
\frac{\alpha(0)}{4a^3}r_{\rm TM}(0)
(\zeta_1^2+2\zeta_1+2)e^{-\zeta_1},
\nonumber
\end{eqnarray}
\noindent
where, in accordance with (\ref{eq13}),
$\zeta_1=4\pi ak_BT/(\hbar c)$, i.e., ${\cal F}_1$ is really exponentially
small. Note that due to (\ref{eq5})
\begin{equation}
r_{\rm TM}(0)=\frac{\varepsilon(0)-1}{\varepsilon(0)+1}
\quad \mbox{and} \quad
r_{\rm TM}(0)=1
\label{eq22}
\end{equation}
\noindent
for dielectric and metallic surfaces, respectively.

By taking the zero-frequency term, $l=0$, in (\ref{eq14}),
one arrives at the nonperturbative free energy of an atom-surface interaction
in the classical limit
\begin{equation}
{\cal F}_0(a,T)=\frac{k_BT}{2}
\ln \left[1-\frac{\alpha(0)}{4a^3}
\int_{0}^{\infty}y^2e^{-y}dy\right]
=\frac{k_BT}{2}
\ln \left[1-\frac{\alpha(0)}{2a^3}\right].
\label{eq23}
\end{equation}
\noindent
In the first perturbation order in the small parameter $\alpha(0)/a^3$
this result is the familiar classical limit of an atom-surface interaction
following from the Lifshitz formula (\ref{eq16}) \cite{9}
\begin{equation}
{\cal F}_0^{\rm per}(a,T)=-\frac{k_BT}{4a^3}\alpha(0).
\label{eq24}
\end{equation}

In a similar way, for the nonperturbative atom-surface force from (\ref{eq15})
we obtain the classical limit
\begin{equation}
{F}_0(a,T)=-\frac{k_BT\alpha(0)}{8a^4}
\frac{\int_{0}^{\infty}y^3e^{-y}dy}{1-\frac{\alpha(0)}{4a^3}
\int_{0}^{\infty}y^2e^{-y}dy}
=-\frac{3k_BT}{4a^4}\alpha(0)
\frac{1}{1-\frac{\alpha(0)}{4a^3}}.
\label{eq25}
\end{equation}
\noindent
In the first perturbation order this leads to
\begin{equation}
{F}_0^{\rm per}(a,T)=-\frac{3k_BT}{4a^4}\alpha(0),
\label{eq26}
\end{equation}
\noindent
i.e., to the same result as
follows \cite{9} from the standard Lifshitz formula (\ref{eq16}).

Now we consider the exact free energy of an atom interacting with an
ideal-metal plane. In this case
\begin{equation}
r_{\rm TM}(i\zeta_l,y)=-r_{\rm TE}(i\zeta_l,y)=1
\label{eq27}
\end{equation}
\noindent
and from (\ref{eq14}) one obtains
\begin{eqnarray}
&&
{\cal F}(a,T)=k_BT\sum_{l=0}^{\infty}{\vphantom{\sum}}^{\prime}
\ln \left[1-\frac{\alpha(i\omega_c\zeta_l)}{4a^3}
\int_{\zeta_l}^{\infty}y^2e^{-y}dy\right]
\nonumber \\
&&
=k_BT\sum_{l=0}^{\infty}{\vphantom{\sum}}^{\prime}
\ln \left[1-\frac{\alpha(i\omega_c\zeta_l)}{4a^3}
(\zeta_l^2+2\zeta_l+2)e^{-\zeta_l}\right].
\label{eq28}
\end{eqnarray}
\noindent
Expanding the logarithm in a power series we present the results in
the following convenient form:
\begin{equation}
{\cal F}(a,T)=-k_BT
\sum_{n=1}^{\infty}\frac{1}{n}
\left(\frac{\alpha(0)}{4a^3}\right)^n
\sum_{l=0}^{\infty}{\vphantom{\sum}}^{\prime}
\left(\frac{\alpha(i\omega_c\zeta_l)}{\alpha(0)}\right)^n
(\zeta_l^2+2\zeta_l+2)^ne^{-n\zeta_l}.
\label{eq29}
\end{equation}
\noindent

The transition to zero temperature is performed by the standard replacement
\begin{equation}
k_BT\sum_{l=0}^{\infty}{\vphantom{\sum}}^{\prime}{\,}\to{\,}
\frac{\hbar c}{4\pi a}\int_{0}^{\infty}d\zeta.
\label{eq30}
\end{equation}
\noindent
This transforms the atom-surface free energy (\ref{eq29}) into the energy
\begin{equation}
{E}(a)=-\frac{\hbar c}{4\pi a}
\sum_{n=1}^{\infty}\frac{1}{n}
\left(\frac{\alpha(0)}{4a^3}\right)^n
\times\int_{0}^{\infty}d\zeta
\left(\frac{\alpha(i\omega_c\zeta)}{\alpha(0)}\right)^n
(\zeta^2+2\zeta+2)^ne^{-n\zeta}.
\label{eq31}
\end{equation}
\noindent
At sufficiently large atom-surface separations considered by Casimir and Polder
\cite{4} one can put $\alpha(i\omega_c\zeta)=\alpha(0)$.
Separating the term with $n=1$ in (\ref{eq31}), we find
\begin{eqnarray}
&&
{E}(a)=-\frac{\hbar c\alpha(0)}{16\pi a^4}\int_{0}^{\infty}d\zeta
(\zeta^2+2\zeta+2)e^{-\zeta}
\label{eq32} \\
&&~~~~
-\frac{\hbar c}{4\pi a}
\sum_{n=2}^{\infty}\frac{1}{n}
\left(\frac{\alpha(0)}{4a^3}\right)^n
\int_{0}^{\infty}d\zeta
(\zeta^2+2\zeta+2)^ne^{-n\zeta}.
\nonumber
\end{eqnarray}
After an integration and changing the index of summation,
one finally obtains
\begin{eqnarray}
&&
{E}(a)=-\frac{3\hbar c\alpha(0)}{8\pi a^4}
-\frac{\hbar c\alpha(0)}{16\pi a^4}
\sum_{k=1}^{\infty}\frac{1}{k+1}
\left(\frac{\alpha(0)}{4a^3}\right)^k
\nonumber\\
&&~~~~~~~
\times
\int_{0}^{\infty}d\zeta
(\zeta^2+2\zeta+2)^{k+1}e^{-(k+1)\zeta}.
\label{eq33}
\end{eqnarray}
\noindent
Note that one power of $\alpha(0)/(4a^3)$ is placed in front of the
summation sign.
The first term on the right-hand side of (\ref{eq33}) is the famous
result by Casimir and Polder \cite{4}, whereas the infinite sum in powers of
the parameter $\alpha(0)/(4a^3)$ presents small nonperturbative correction to
this result.

\section{Conclusions and discussion}

In the foregoing, we have considered the nonperturbative theory
describing the interaction of a ground-state atom
with a real-material surface at any temperature in
thermal equilibrium with the environment. After
presenting several main equations of this theory, we
have made a transition to the Matsubara representation
which allowed detailed comparison between the exact
free energies and forces of atom-surface interaction
and that ones calculated perturbatively using the
standard Lifshitz formulas. The latter have been
obtained from the nonperturbative results as the first
perturbation order in the small parameter equal to the
dynamic atomic polarizability divided by the third
power of an atom-surface separation.

Computations of the atom-surface free energies and forces
have been performed by using both the exact and
perturbative theories for the atoms of He${}^{\ast}$ and Na
interacting with an Au wall at room temperature. In
so doing the highly accurate dynamic atomic
polarizabilities and the tabulated optical data for
the complex index of refraction of Au have been used.
It was shown that the maximum deviations between the
predictions of both theories are reached at the
minimum separation considered ($a$ = 0.8\,nm). For an
atom of He${}^{\ast}$, the maximum relative deviations between
the two theories in the computed free energy and force
are equal to approximately 1.1\% and 2.3\%,
respectively. For an atom of Na, the maximum
deviations reach 0.6\% for the free energy and 1.25\%
for the force. It was concluded that the predicted
deviations from the results of perturbative theory are
experimentally observable.

Finally, we have derived simple analytic expressions for
the classical free energy and force of an atom-surface
interaction in the framework of nonperturbative theory.
In the lowest order of the small parameter mentioned
above, the obtained expressions coincide with the
well known results following from the standard Lifshitz
formulas. The free energy of an atom interacting with
an ideal-metal wall has also been found using the
nonperturbative theory. In the lowest order of the same small
parameter it coincides with the commonly known Casimir
and Polder perturbative result.

The exact expressions for an atom-surface interaction
considered in this paper are interesting not only
from the fundamental point of view, but may find
applications for the interpretation of experiments on
quantum reflection and in the theory of the van der Waals
adsorption. The latter may need the more precise interaction
potentials at the shortest atom-surface separation than
that ones obtained perturbatively using the standard approach.

\ack{The work of V.M.M. was partially supported by the Russian
Government
Program of Competitive Growth of Kazan Federal University.}

\newpage
\end{document}